\documentclass[12pt]{article}
\begin{document}
\vspace*{3cm}
\begin{center}
\textbf{\LARGE Nuclear collective processes study \\  \smallskip
with attosecond laser pulses}

\vspace{2cm}

{\large Janina Marciak-Kozlowska
\bigskip

 and

\bigskip
Miroslaw Kozlowski}\footnote{Corresponding author, e-mail: MiroslawKozlowski@aster.pl,\\ http://www.fuw.edu.pl/$\sim$mirkoz}

\vspace{2cm}

Institute of Electron Technology, Al. Lotnik\'{o}w 32/46, 02-Warsaw,
Poland

\end{center}

\newpage
\begin{abstract}
In this paper the possibility of the excitation of collective
nuclear motion with attosecond laser pulses is investigated.
Following the results of our earlier results (\textit{Lasers in Engineering}, \textbf{11}
(2001), p.~259) the hyperbolic heat transport for nuclear
matter is formulated and solved. It is shown that in the vicinity
of the 30~MeV excitation energy the recollided electrons
can excite the giant collective motions -- thermal wave inside
the nuclei.

\end{abstract}

\newpage
\section{Introduction}
With attosecond lasers (1{\nobreakspace}as = 10$^{-18}${\nobreakspace}s) physicist
and engineers are now close to controlling the motion of electrons
on a timescale that is substantially shorter than the oscillation
period of visible light. It is also possible with attosecond
laser pulse to rip an electron wave-packet from the core of an
atom and set it free with similar temporal precision.

Recently a laser configuration in which attosecond electron wave
packets are ionized and accelerated to multi-MeV energies, was
proposed~\cite{1}. This technique opens an avenue towards imaging
attosecond dynamics of nuclear processes.

High laser intensity atomic and molecular physics is dominated
by the recollision between and ionized electron and its parent
ion. The electron is ionized near the peak of laser field, accelerated
away from the ion and driven back to its parent ion once the
field direction reverses. Recollision leads to nonsequential
double ionization, high harmonic generation, and attosecond extreme
ultraviolet and electron pulses~\cite{1}.

In contrast the extension of laser induced recollision physics
to relativistic energies is a long standing issue. The solution
to this problem was achieved in paper~\cite{1}. The authors
of the paper~\cite{1} show that the Lorentz force (which prevents
recollision for relativistic electrons) is eliminated for two
counterpropagating, equally handed, circularly polarized beams
through the whole focal volume as long as the laser pulses are
sufficiently long. In this configuration, the recollision energy
is limited only by the maximum achievable laser intensity (currently
\ensuremath{\sim}{\nobreakspace}10$^{23}$ W/cm$^{3}$).

The method presented in paper~\cite{1} will alter the physics
anywhere relativistic laser fields interact with electrons. For
example the attosecond laser pulses can resolve nuclear dynamics.
In nuclear physics it opens novel possibilities in the study
of decay and damping of nuclear dynamical processes such as giant
resonances. It reveals information on fundamental nuclear properties,
such as dissipation and viscosity in nuclei.

\section{Relativistic hyperbolic heat transport equation for nuclear processes}

In paper~\cite{3} relativistic hyperbolic transport equation{\nobreakspace}(RHT)
was formulated:
    \begin{equation}
    \frac{1}{v^{2} } \frac{\partial ^{2} T}{\partial t^{2} } +\frac{m_{0}
    \gamma }{\hbar } \frac{\partial T}{\partial t} =\nabla ^{2} T.
    \label{eq1}
    \end{equation}
In equation~(\ref{eq1}) $v$ is the velocity of heat waves,
$m_{0} $
 is the mass of heat carrier (nucleon) and
$\gamma $
-- the Lorentz factor, $\gamma =\left( 1-\frac{v^{2} }{c^{2} } \right) ^{-\frac{1}{2} } .$
 As was shown in paper~\cite{1} the heat energy (\textit{heaton
temperature})
$T_{h} $
 can be defined as follows:
    \begin{equation}
    T_{h} =m_{0} \gamma v^{2}.\label{eq2}
    \end{equation}
Considering that $v$, thermal wave velocity equals~\cite{1}
    \begin{equation}
    v=\alpha c.\label{eq3}
    \end{equation}
where $\alpha $
 is the coupling constant for the interactions which generate
the \textit{thermal wave} ($\alpha =0.15$
 for strong force) \textit{heaton temperature} equals
    \begin{equation}
    T_{h} =\frac{m_{0} \alpha ^{2} c^{2} }{\sqrt{1-\alpha ^{2} } }.\label{eq4}
    \end{equation}
From formula~(\ref{eq4}) one concludes that \textit{heaton temperature}
is the linear function of the mass{\nobreakspace}$m_{0} $
of the heat carrier. It is quite interesting to observe that
the proportionality of $T_{h} $
 and heat carrier mass $m_{0} $
 was the first time observed in ultrahigh energy heavy ion reactions
measured at CERN~\cite{4}. In paper~\cite{4} it was shown
that temperature of pions, kaons and protons produced in Pb+Pb,
S+S reactions are proportional to the mass of particles. Recently
at Rutherford Appleton Laboratory (RAL) the VULCAN laser was
used to produce the elementary particles: electrons and pions~\cite{5}.

In the present paper the forced relativistic heat transport equation
will be studied and solved. In paper~\cite{6} the damped thermal
wave equation was developed:
    \begin{equation}
    \frac{1}{v^{2} } \frac{\partial ^{2} T}{\partial t^{2} } +\frac{m}{\hbar}
    \frac{\partial T}{\partial t} +\frac{2Vm}{\hbar ^{2} } T-\nabla ^{2} T=0.\label{eq5}
    \end{equation}
The relativistic generalization of equation~(\ref{eq5}) is quite
obvious:
    \begin{equation}
    \frac{1}{v^{2} } \frac{\partial ^{2} T}{\partial t^{2} } +\frac{m_{0}
    \gamma }{\hbar } \frac{\partial T}{\partial t} +\frac{2Vm_{0} \gamma}{\hbar ^{2} }
    T-\nabla ^{2} T=0.\label{eq6}
    \end{equation}
It is worthwhile to note that in order to obtain nonrelativistic
equation we put $\gamma =1$.

The motion of charged nucleons in the nucleus is equivalent to
the flow of an electric current in a loop of wire. With attosecond
laser pulses we will be able to influence the current in the
nucleon ``wire''. This opens quite new perspective for the attosecond
nuclear physics. The new equation~(\ref{eq6}) is the natural candidate
for the master equation which can be used to the description
of heat transport in nuclear matter.

When external force is present $F(x,t)$ the forced damped
heat transport is obtained instead of equation~(\ref{eq6}) (in
one dimensional case):
    \begin{equation}
    \frac{1}{v^{2} } \frac{\partial ^{2} T}{\partial t^{2} } +\frac{m_{0}
    \gamma }{\hbar } \frac{\partial T}{\partial t} +\frac{2Vm_{0} \gamma}{\hbar ^{2} }
    T-\frac{\partial ^{2} T}{\partial x^{2} } =F(x,t).\label{eq7}
    \end{equation}
The hyperbolic relativistic quantum heat transport equation (RQHT),
formula~(\ref{eq7}), describes the forced motion of heat carriers
which undergo the scatterings
$\left( \frac{m_{0} \gamma }{\hbar } \frac{\partial T}{\partial t}
\;{\rm term}\right) $
and are influenced by potential $\left(\frac{2Vm_{0} \gamma }{\hbar ^{2} } T\;{\rm term}\right)$.

The solution of equation can be written as
    \begin{equation}
  T(x,t)=e^{-\frac{t}{2\tau } } u(x,t),\label{eq8}
    \end{equation}
where $\tau =\frac{\hbar }{\left( mv^{2} \right) } $
is the relaxation time. After substituting formula~(\ref{eq8})
to the equation~(\ref{eq7}) we obtain new equation
    \begin{equation}
    \frac{1}{v^{2} } \frac{\partial ^{2} u}{\partial t^{2} } -\frac{\partial^{2} u}
    {\partial x^{2} } +qu(x,t)=e^{\frac{t}{2\tau } } F(x,t),\label{eq9}
    \end{equation}
and
    \begin{eqnarray}
    q&=&\frac{2Vm}{\hbar ^{2} } -\left( \frac{mv}{2\hbar } \right) ^{2} ,\label{eq10}\\
    m&=&m_{0} \gamma.\nonumber
    \end{eqnarray}
Equation~(\ref{eq9}) can be written as:
    \begin{equation}
    \frac{\partial ^{2} u}{\partial t^{2} } -v^{2} \frac{\partial ^{2}
    u}{\partial x^{2} } +qv^{2} u(x,t)=G(x,t),\label{eq11}
    \end{equation}
where
    $$
    G(x,t)=v^{2} e^{\frac{t}{2\tau } } F(x,t).
    $$

When $q>0$ equation~(\ref{eq11}) is the forced Klein-Gordon (K-G) equation.
The solution of the forced Klein-Gordon equation for the initial
conditions:
    \begin{equation}
    u(x,0)=f(x),\quad u_{t} (x,0)=g(z)\label{eq12}
    \end{equation}
has the form~\cite{4}:
    \begin{eqnarray}
    u(x,t)&=&\frac{f(x-vt)+f(x+vt)}{2}\label{eq13}\\
    &&\mbox{}+\frac{1}{2v} \int\limits_{x-vt}^{x+vt}g(\zeta  )J_{0}
    \left[q\sqrt{v^{2} t^{2}-(x-\zeta )^{2} } \right] d\zeta\nonumber \\
    &&\mbox{}-\frac{\sqrt{q} vt}{2} \int\limits_{x-vt}^{x+vt}f(\zeta )\frac{J_{1}
    \left[ q\sqrt{v^{2} t^{2} -(x-\zeta )^{2} } \right] }{\sqrt{v^{2} t^{2}
    -(x-\zeta )^{2} } } d\zeta\nonumber \\
    &&\mbox{}+\frac{1}{2v}
    \int\limits_{0}^{t'}\int\limits_{x-v(t-t')}^{x+v(t-t')}G(\zeta ,t')J_{0}
    \left[ q\sqrt{v^{2} (t-t')^{2} -(x-\zeta )^{2} } \right] d\zeta dt'.\nonumber
    \end{eqnarray}
When $q<0$  equation~(\ref{eq11}) is the forced modified Heaviside (telegraph)
equation with the solution~\cite{5}
    \begin{eqnarray}
    u(x,t)&=&\mbox{}\frac{f(x-vt)+f(x+vt)}{2}\label{eq14} \\
    &&\mbox{}+\frac{1}{2v} \int\limits_{x-vt}^{x+vt}g(\zeta )J_{0} \left[
    -q\sqrt{v^{2} t^{2} -(x-\zeta )^{2} } \right] d\zeta \nonumber \\
    &&\mbox{}+\frac{\sqrt{-q} vt}{2} \int\limits_{x-vt}^{x+vt}f(\zeta )\frac{J_{1}
    \left[ -q\sqrt{v^{2} t^{2} -(x-\zeta )^{2} } \right] }{\sqrt{v^{2} t^{2}
    -(x-\zeta )^{2} } } d\zeta\nonumber \\
    &&\mbox{}+\frac{1}{2v}
    \int\limits_{0}^{t'}\int\limits_{x-v(t-t')}^{x+v(t-t')}G(\zeta ,t')J_{0}
    \left[ -q\sqrt{v^{2} (t-t')^{2} -(x-\zeta )^{2} } \right] d\zeta dt'.\nonumber
    \end{eqnarray}
When $q=0$  equation~(\ref{eq11}) is the forced thermal equation~\cite{7}
    \begin{equation}
    \frac{\partial ^{2} u}{\partial t^{2} } -v^{2} \frac{\partial ^{2}
    u}{\partial x^{2} } =G(x,t).\label{eq15}
    \end{equation}
On the other hand one can say that equation~(\ref{eq15}) is the
distortionless hyperbolic equation. The condition $q=0$
 can be rewrite as:
    \begin{equation}
    V\tau =\frac{\hbar }{8}.\label{eq16}
    \end{equation}
The equation~(\ref{eq16}) is the analogous to the Heisenberg uncertainty
relations. Considering formula~(\ref{eq2}) equation~(\ref{eq16})
can be written as:
    \begin{equation}
    V=\frac{T_{h} }{8} ,\quad V<T_{h}.\label{eq17}
    \end{equation}
One can say that the distortionless waves can be generated only
if $T_{h} >V$. For $T_{h} <V$, i.e. when the ``Heisenberg rule'' is broken, the shape of the
thermal waves is changed.

We consider the initial and boundary value problem for the inhomogenous
thermal wave equation in semi-infinite interval~\cite{7}: that
is
    \begin{equation}
    \frac{\partial ^{2} u}{\partial t^{2} } -v^{2} \frac{\partial ^{2}
    u}{\partial x^{2} } =G(x,t),\quad 0<x<\infty ,\quad t>0,\label{eq18}
    \end{equation}
with initial condition:
    $$
    u(x,0)=f(x),\quad \frac{\partial u(x,0)}{\partial t} =g(x),\quad
    0<x<\infty ,
    $$
and boundary condition
    \begin{equation}
    au(0,t)-b\frac{\partial u(0,t)}{\partial x} =B(t),\quad t>0,\label{eq19}
    \end{equation}
where $a\geq 0,\;b\geq 0,\;a+b>0$
 (with $a$ and $b$ both equal to constants) and $F$, $f$, $g$ and $B$ are given functions. The solution of equation~(\ref{eq18}) is of the form~\cite{4}
    \begin{eqnarray}\label{eq20}
    u(x,t)&=&\mbox{}\frac{1}{2} \left[ f(x-vt)+f(x+vt)\right] +\frac{1}{2v}
    \int\limits_{x-vt}^{x+vt}g(s)ds\\
    &&\mbox{}+\frac{1}{2v}
    \int\limits_{0}^{t}\int\limits_{x-v(t-t')}^{x+v(t-t')}F(s,t')dsdt'. \nonumber
    \end{eqnarray}
In the special case where $f=g=F=0$
we obtain the following solution of the initial and boundary
value problem~(\ref{eq19}), (\ref{eq20}):
    \begin{eqnarray}\label{eq21}
    u(x,t)=\left\{\begin{array}{lcl}
    0, &\quad& x>vt, \\
\displaystyle\frac{v}{b}\int_0^{t-\frac{x}{v}}
\exp\left[\frac{va}{b}\left(y-t+\frac{x}{v}\right)\right]B(t)dt, &\quad& 0<x<vt\end{array}\right.
    \end{eqnarray}
if $b\neq 0.$
 If $a=0$  and $b=1$, we have:
    $$
    u(x,t)=\left\{
    \begin{array}{lcl}
    0, &\quad& x>vt, \\
    \displaystyle v\int_0^{t-\frac{x}{v}}B(y)dy &\quad& 0<x<vt
    \end{array}
    \right.
    $$

It can be concluded that the boundary condition{\nobreakspace}(19) gives
rise to a wave of the form $K\left( t-\frac{x}{v} \right) $
 that travels to the right with speed \textit{v}. For this reason
the foregoing problem is often referred to as a \textit{signaling
problem} for the thermal waves.\\

\section{Attosecond laser electronuclear spectroscopy}
As was shown in paragraph{\nobreakspace}2 the master equation for collective
(thermal wave) motion of nucleons has the form
    \begin{equation}
    \frac{\partial ^{2} u}{\partial t^{2} } -v^{2} \frac{\partial ^{2}u}
    {\partial x^{2} }+qv^{2} u(x,t)=G(x,t).\label{eq22}
    \end{equation}
When we are looking for undistorted motion, $q\rightarrow 0$
 i.e.:
    $$
    \frac{2Vm}{\hbar ^{2} } -\left( \frac{mv}{2\hbar } \right) ^{2}\rightarrow 0.
    $$
For $q=0$  one obtain
    \begin{equation}
    V\tau =\frac{\hbar }{8}.\label{eq23}
    \end{equation}
For nuclear collective motion $\tau =\frac{\hbar }{mv^{2} } $, $m$ is the nucleon mass and
$v=\alpha _{s} c$, where $\alpha _{s} =0.15$~\cite{8} is the strong coupling constant. With the $m=981\frac{MeV}{c^{2} } $  one obtains from formula~(\ref{eq23})
    \begin{equation}
    V=\frac{\hbar }{8\tau } \approx 30 {\rm MeV}.\label{eq24}
    \end{equation}

As it is well known the 30{\nobreakspace}MeV energy range is the location
of giant dipole resonances in nuclear matter~\cite{9}.

As was shown in paper~\cite{2} for attosecond laser induced
recollision of electrons with ions the energy of the electrons
is of the order of 30{\nobreakspace}MeV for
$Z\approx 20$
 (where $Z$ is the atomic number of the ion). It means that
with attosecond laser pulses the electronuclear giant resonances
can be excited and investigated.
\section*{Conclusion}
In this paper the interaction of the attosecond laser pulses
with atoms was investigated. It is shown that recollided (after
attoseconds pulse excitations) electrons can excite the thermal
wave -- giant resonances in nuclear matter. These electronuclear
reactions can be investigated with hyperbolic nuclear thermal
equation~(\ref{eq18}).

\end{document}